\documentclass[%
 reprint,
superscriptaddress,
groupedaddress,
 amsmath,amssymb,
aps,
prb,
]{revtex4-2}
\usepackage{epsfig} % pour g´erer les images
\usepackage{float} % pour le placement des figure
\usepackage{balance}
\usepackage{color}
\usepackage{graphicx}% Include figure files
\usepackage{dcolumn}% Align table columns on decimal point
\usepackage{bm}% bold math
\usepackage{hyperref}
\usepackage[toc,page]{appendix}

\hypersetup{
colorlinks=true,
linkcolor=blue,
filecolor=red,
urlcolor=blue,
citecolor=blue
}
\usepackage[mathlines]{lineno}% Enable numbering of text and display math
\relax % Commence numbering lines
\usepackage{physics}
\usepackage{here}
\usepackage[english]{babel}
\usepackage[utf8]{inputenc} 
\usepackage{subcaption}
\usepackage{lipsum}
\usepackage{textcomp}
\usepackage[T1]{fontenc} % pour les font postscript
\usepackage[cyr]{aeguill} % Police vectorielle TrueType, guillemets fran¸cais
\usepackage{epsfig} % pour g´erer les images
\usepackage[%showframe,%Uncomment any one of the following lines to test 
]{geometry}
\DeclareUnicodeCharacter{1EF3}{\`y}
\DeclareUnicodeCharacter{2032}{'}
\DeclareUnicodeCharacter{0301}{`}
\DeclareUnicodeCharacter{0308}{"}
\begin{document}
\captionsetup[figure]{labelfont={bf},name={Fig.},labelsep=period}
\title{Quantum beats of coherent 1s 2s excitons in two dimensional transition metal dichalcogenides}

\author{{ Nizar Chaouachi$^1$, Sihem Jaziri$^{1,2}$}\\
{\small \em $^1$Facult\'{e} des Sciences de Tunis, Laboratoire de Physique de la Mati\`{e}re Condens\'{e}e,
D\'{e}partement de Physique, Université Tunis el Manar, Campus Universitaire 2092 Tunis, Tunisia\\
$^2$ Facult\'{e} des Sciences de Bizerte, Laboratoire de Physique des Mat\'{e}riaux Structure et Propri\'{e}t\'{e}s , Universit\'{e} de Carthage,7021 Jarzouna, Tunisia
}}
\date{\today}
\begin{abstract}
\begin{center}
\underline{Abstract}
\end{center}
Motivated by recent experimental measurement of the intrinsic excitonic wave-function in
2D Transition-metal dichalcogenides (TMDs) by angle-resolved photoemission spectroscopy
(ARPES), we developed a theoretical study to resolve some characteristics of these excitons and
some of the many open issues in these systems. The system is assumed to be embedded in an
environment with average dielectric constant $\kappa$, below which electrostatic interactions in the
corresponding TMD layer are screened. We adopt the long range approximation, which gives the
electron-hole interaction in the Rytova-Keldysh form. Latter allows understanding the role of
screening in TMDs structures. The bound state 1s, 2s… energy eigenvalues for the two-dimensional
are reformulated in momentum space leads to an integral form of the Wannier equation. The
eigenfunctions are then expanded in terms of spherical harmonics. To evaluate the dynamic of the
angle-resolved photoemission spectrum arising from the dissociation of excitons given their steady-
states 1s, 2s.. expressions, we follow the semi perturbative theoretical description developed by
previous calculations. We discuss the dielectric environment effect on the dispersive features of the
spectrum for different 1s, 2s,… exciton distributions. Quantum beat signatures in photoemission
intensity demonstrate coherent coupling between 1s and 2s excitons. the beating contribution due
to excitonic coherence is also discussed.The periodic oscillations arising from coherent superposition states, quantum beats,  enable exploration of novel coherent phenomena.\\
\end{abstract}
\pacs{Valid PACS appear here}% PACS, the Physics and Astronomy
                             % Classification Scheme.
\maketitle
\section{Introduction}
The exciton interaction with the electromagnetic field may lead to significant phenomenas. Moreover the bosonic character of excitons allow us to reveal Bose-Einstein-Condensate-like phenomena can be described as coherence of excitons \cite{coh1}\cite{coh2}. Using femto-second lasers, it is possible to create two coherent excitons with two different energy levels simultaneously\cite{quab1}. The oscillatory behavior of the light intensity emitted by the two-levels excited system is defined as quantum beats\cite{quab2}. However only limited observations were reported For the quantum beats in excitonic complexes in Transition Metal Dichalcogenides (TMDs) \cite{beat1}\cite{beat2}\cite{beat3}\cite{beat4}\cite{beat5}\cite{beat6}.\\
Two-dimensional TMDs are  materials composed from one layer of transition metal (M,W$\cdots$) between two layers of chalcogenides (S,Se,Te) \cite{1.4}\cite{1.5}\cite{1.6}\cite{1.7}\cite{1.8}\cite{1.9}\cite{1.10}. TMDs family have exhibited many intrinsic electronic and optical properties like semiconductors with direct band gap in visible spectrum, metals, superconductors, semi-metals Weyl semi-metals, Mott-insultors and important spin orbit coupling \cite{1.9}\cite{1.10}\cite{1.11}\cite{1.12}\cite{1.121}\cite{1.123}\cite{1.124}.\\ All these properties have enlarged the application field of these materials from ultra-thin transistor, photovoltaic panels, energy stocking to new domains like spintronics, Valleytronics, plamonics\cite{1.9}\cite{1.13}\cite{1.15}\cite{1.16}\cite{1.17}\cite{1.18}\cite{1.19}. The optical response in TMDs semi-conductors are dominated by excitonic effects. Originating from strong reduced dielectric screening, two-dimensional transition metal dichalcogenides exhibit strongly bound excitons with an energy overcomes the room temperature. Thus it is possible to manipulate these quasi-particles \cite{ex1}\cite{ex2}\cite{ex3}\cite{ex4}\cite{ex5}\cite{ex6}. Recent studies have shown  Rydberg states  in monolayer $\mathrm{MoS_2}$ \cite{38}\cite{41mag} which make it an ideal platform for studying coherent phenomena. Motivated by recent experimental measurement of the intrinsic excitonic wave-function as well as  by  time and angle-resolved photoemission spectroscopy (tr-ARPES) \cite{15}\cite{15.1}, we develop a theoretical study to resolve some characteristics of these excitons and some of the many open issues in these systems.\\
Time resolved ARPES have provided a plethora of information about non-equilibrium states because of its controllable environment as well as the high resolution on energetic, angular and time scales using ultra-short lasers pulses  \cite{4}\cite{5}. This technique have demonstrated a success in investigating strongly correlated systems \cite{6}, exotic phases such as Weyl semi-metals \cite{7}, superconductors \cite{8}, charge density wave \cite{9} and topological insulators \cite{10}. \\
Among its advantages is that it maps a large energy spectrum as well as the momentum. Hence it is possible to observe various excitons including A, B excitons, dark, spin and momentum forbidden excitons along with excitonic insulators \cite{10}\cite{11}\cite{12}\cite{13}\cite{14}.
In this article we first describe excitons  in 2D materials and the environment effect on the density of probability and the energy using Wannier equation. The second section is dedicated to the theoretical formalism of the time resolved photoemission spectroscopy resulted from excitons in 2D semi-conductors with direct gap based on Freerick et al. \cite{16} which provides a description of time resolved photoemission spectroscopy within non-equilibrium states, and A.F Kemper et al.  \cite{17}\cite{18} for theoretical description of photoemission of exciton in bulk materials with indirect gap. The results of photoemission spectra and the effect of environment for excitons as well as the quantum beats are discussed and compared to other results in the third part.
\section{Exciton in Transition Metal Dichakcogenides}
We consider a model two-band 2D TMD semiconductor within the effective mass approximation. Note that we take into account the conduction band minima and the valence band maxima at the K or K’ valley, which are crucial for the formation of momentum  exciton states. \\

A monolayer is always embedded between two media. In this case, it is more convenient to describe the screened interaction by Rytova-Keldysh potential $V_{RK}(r)= \frac{-e^2}{4\pi\varepsilon_0\kappa r_s}\{H_0(\frac{r}{r_s})-Y_0(\frac{r}{r_s})\}$, where $H_0(x)$ and $Y_0(x)$ are the zero order Sturve and Newmann functions, r=$\|\vec{r}_e-\vec{r}_h\| $ is the in plane relative distance between the electron and hole, $r_s=\frac{\varepsilon_r d}{2\kappa}$ is the effective screening length, $\kappa=\frac{\varepsilon_t +\varepsilon_b}{2}$ is the average dielectric constant of the two media surrounding the monolayer where $\varepsilon_b$ is the dielectric constant of the bottom medium and $\varepsilon_t$ is the dielectric constant of the top medium, d is the monolayer width and $\varepsilon_r$ is its dielectric constant \cite{19.1}\cite{20}\cite{21}\cite{22}

\begin{figure}[H]
\centering\
\includegraphics[scale=0.3]{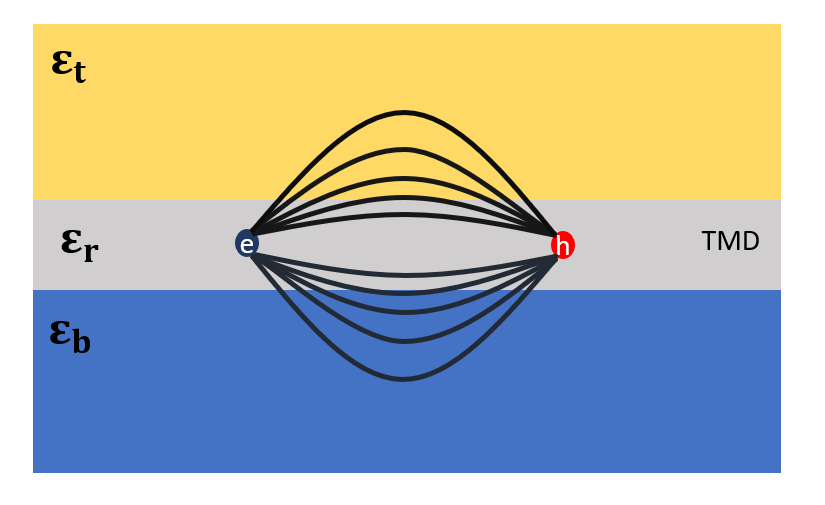}
\caption{Schematic illustration of field lines between electron and hole in the monolayer  embedded between two media}
\end{figure}

To obtain access to the wave-functions and eigenvalues of excitons, we solve the eigenvalue problem of the following Hamiltonian  \cite{23}\cite{ex5} :
\begin{equation}
H_{X_{Q,k}}=\frac{\hbar^2Q^2}{2M}+\frac{\hbar^2k^2}{2\mu}+V_{RK}(r)+E_{g}
\end{equation}
Here $\vec{Q}=\vec{k}_e-\vec{k}_h$ is the center of mass momentum, $\vec{k}=\frac{m_{h}\vec{k}_e+m_{e}\vec{k}_h}{m_{h}+m_{e}}$ is the relative momentum , $M=m_e+m_h$ is the total mass , $\mu = \frac{m_{e}m_{h}}{m_{h}+m_{e}}$ is the reduced mass of the exciton and $E_{g}$ the band gap energy of the material.
 $\frac{\hbar^2k^2}{2\mu}+V_{RK}(r)$ defines the in-plane relative motion hamiltonian $H_{rel}$. Hence we write the momentum space Wannier equation of relative motion in the unit of Bohr radius $a_X=\frac{\varepsilon_r \hbar^2}{\mu e^2}$ :
\begin{eqnarray}
& & q^2 \Phi_{nm}(\vec{q})-\frac{1}{\pi}\int \frac{\Phi_{nm}(\vec{q}')d\vec{q}}{|\vec{q}-\vec{q}'|}\nonumber\\
& &+\frac{1}{\pi}\int \frac{\Phi_{nm}(\vec{q}')d\vec{q}}{|\vec{q}-\vec{q}'|+\vec{q}_s}=E^{*}_{nm}\Phi_{nm}(\vec{q})
\end{eqnarray} 
$\vec{q}=\vec{k}a_X$ is the dimensionless relative wave vector and $q_s=\frac{a_X}{r_s}$. Here the first two terms representing $H_{hyd}$ which is the effective 2D hydrogenic-like Hamiltonian but we replace $\varepsilon_r$ by $2\kappa$. Therefore, the modified eigenvalues and the effective Bohr radius  are $E^{hyd}_{nm}=-\frac{\varepsilon_{r}^{2}}{\kappa^2} \frac{R_X}{(n-1/2)^2}$ knowing that $R_X=\frac{\mu e^4}{2\varepsilon_{r}^{2}\hbar^2}$ is the effective Rydberg and $a^{*}_{X}(n)=\frac{\kappa}{\varepsilon_r}a_X(n-1/2)$ where \textit{n}=1,2,3 \ldots is the principal quantum number, we notice that the new eigenvalues depend explicitly on the two dielectric constants $\varepsilon_t$ and $\varepsilon_b$  .The corresponding eigenstate are described in cylindrical coordinates \cite{25} :
\begin{eqnarray}
& &\varphi_{nm}(q,\phi)=(-\mathit{i})^m\sqrt{2\pi}(\frac{\frac{2}{n-1/2}}{q^2+(\frac{1}{n-1/2})^2})^\frac{3}{2}\nonumber\\
& &\times P^{|m|}_{n-1}(\frac{q^2-(\frac{1}{n-1/2})^2}{q^2+(\frac{1}{n-1/2})^2})e^{-\mathit{i}m\phi}
\end{eqnarray}
Marking that $P^{|m|}_{\textit{n}-1}(x)$ is the associated Legendre Polynomial, $\phi$ is the polar angle \textit{m}= -(\textit{n}-1)\ldots \textit{n}-1 is the magnetic number but for this work we will consider only the states $\lambda = ns$ i.e m=0.\\
 To find the energies of relative exciton motion  by the screened potential, we numerically solve the eigenvalue problem equation (2) by direct diagonalization over the basis of eigenstates of the 2D hydrogen-like exciton expressed as \cite{26} :
\begin{equation}
\Phi_{\lambda}(q)=\sum_{\lambda '} c_{\lambda '} \varphi_{\lambda '}(q)
\label{eqphi}
\end{equation} 
The coefficients $c_\lambda '$ are real numbers for ns states obtained by the diagonalization of the matrix. 
Accordingly the exciton state with a generic finite center of mass momentum $\vec{Q}$ is \cite{23}\cite{27} :
\begin{eqnarray}
 \ket{X_{\lambda,K}(\vec{Q})}=& &\frac{1}{\sqrt{A}}\sum_{\vec{k}}\Phi^{*}_{\lambda} (\vec{k})a^{\dag}_{c}(\vec{K}+\vec{k}+\frac{m_e}{M}\vec{Q})\nonumber\\& & \times a_{v}(\vec{K}+\vec{k}-\frac{m_h}{M}\vec{Q})\ket{0}
 \end{eqnarray}
 Where A is the ML area , $a^{\dag}_{c}(\vec{K}+\vec{k}+\frac{m_e}{M}\vec{Q})/a_{v}(\vec{K}+\vec{k}-\frac{m_h}{M}\vec{Q})$ create an electron/hole in the conduction/valence band within the valley K and $\ket{0}$ is the ground state with completely filled valence band. This state have the following energy :
\begin{equation}
E_{X_{\lambda,\vec{Q}}}=\frac{\hbar^2Q^2}{2M}+E_{\lambda}+E_g
\label{energies}
\end{equation}
noting that $E_{\lambda}$ is the relative motion energy for the $\lambda$ state derived from the solutions of equation (2) and $\frac{\hbar^2Q^2}{2M}$ is the exciton center of mass kinetic energy and $E_G$ is the gap energy of the MoS$_2$ ML.
\section{Theoretical formalism of two photons Time and Angle Resolved PhotoEmission Spectroscopy}
To appraise excitonic spectrum we use many body hamiltonians. Foremost we suppose that all charge carriers have same spin direction \cite{28}. Also the exciton density is assumed to be low in order to neglect exciton-exciton interactions \cite{29}. The process is described by two photons where the first photon correspond to the pump photon with Visible near Infra-red  energy. Accordingly this photon leads to creation of excitons in the K valley ($\frac{4\pi}{3a}\vec{k}_x$) where a is the lattice parameter)  which will couple with the electromagnetic field expressed by the following hamiltonian :
\begin{equation}
H^{(1)}(t)=H_X+E_0+H_{vc}(t)
\end{equation}
$H_X=\sum_\lambda E_{X_\lambda} X^{\dag}_{\lambda,\vec{K}}X_{\lambda,\vec{K}}$ describes the exciton energy levels  obtained from equation \ref{energies} corresponding to the state $\lambda$  with zero center of mass $\ket{X_{\lambda,\vec{K}}}=\frac{1}{\sqrt{A}}\sum_{\vec{k}}\phi^{*}_{\lambda} (\vec{k}+\vec{K})a^{\dag}_{c}(\vec{K}+\vec{k})a_{v}(\vec{K}+\vec{k})\ket{0}$. $E_0$ is the energy of the ground state $\ket{0}$ which portrays the unbound electrons in the valence band. The excitons with $\vec{Q}=\vec{0}$  can couple directly with photons and lead to optical absorption. Moreover, the $H_{vc}(t)$ describes the exciton-photon coupling where we place ourselves in Coulomb gauge i.e $\vec{\nabla}\cdot \vec{A}=0$, which implies that the potential vector $\vec{A}$ is only restricted only to his in-plane component. We consider the dipole approximation since our pump photon is in visible domain i.e the wave length is larger than the distance between the charge carriers $r$ ($\lambda_{vis} \gg r$). Thus $H_{vc}(t)$ can be expressed as :
\begin{equation}
H_{vc}=-\sum_\lambda\vec{\epsilon}_{vc}(t)\vec{p}_\lambda (t)
\end{equation}
Here $\vec{\epsilon}_{vc}(t)=\vec{\epsilon}_0e^{\frac{-(t-t_c)^2}{2\sigma_{c}^{2}}}\cos (\Omega(t-t_c))$ is the classical electromagnetic pump pulse centred in $t_c$ with Gaussian profile, $\Omega$ is the frequency in the visible light, $\sigma_c$ is the temporal pulse width. Moreover  $\vec{p}_\lambda (t)=\vec{d}_\lambda (X_{\lambda}^{\dag}e^{\mathit{i}\frac{E_{X_\lambda}}{\hbar}t}+X_{\lambda}e^{-\mathit{i}\frac{E_{X_\lambda}}{\hbar}t})$ is the electric dipole operator \cite{29.1} and $d_\lambda$ is the transition matrix element.  \\
After certain time delay from creating the non-equilibrium state, the probe photon with energy higher than the material's work function \cite{4}\cite{33} will be absorbed by the bound electron. the many body Hamiltonian of the photoemitted electron is expressed as :
\begin{equation}
H^{(2)}(t)=H_f+H_{cf}(t)
\end{equation}
where $H_f=\sum_{\vec{k}_e}(\frac{\hbar^2 k_{e}^{2}}{2m_e}+ W)a_{f}^{\dag}(\vec{k}_e)a_f(\vec{k}_e)$ is the electron kinetic energy after being ejected from the solid with momentum $\vec{k}_e$, $a_{f}^{\dag}(\vec{k}_e)/a_f(\vec{k}_e)$ is the creation/annihilation operator of free electron. $ W$ denotes the work function of the monolayer.  Additionally, $H^{2}_{int}=\sum_{\vec{k}_e,\vec{k'}}\hbar\Omega_{R}^{cf}(t)a_{f}^{\dag}(\vec{k}_e)a_c(\vec{k'}+\vec{K})+h.c$ will annihilate an electron from non equilibrium state and create it in  the vacuum.\\ $\Omega_{R}^{cf}(t)=\frac{d^{cf}\epsilon_{1}}{\hbar}e^{\mathit{i}\omega_{ph}t}e^{-\frac{(t-t_p)^2}{2\sigma_{p}^{2}}}$ is time dependent Rabi frequency with low magnitude of the electromagnetic field $\epsilon_{1}$ and $d^{cf}$ is the transition matrix element, $\hbar\omega_{ph}$ is the photon energy in the Xrays-Ultra-violet domain \cite{5},  $\sigma_p$ is the probe pulse temporal width, $t_p$ is the moment when the pump pulse is maximal and $d^{cf}$ is the dipole transition element which also assumed to be momentum independent.\\ 

Following the evolution of the system described in the Appendix A and B bellow, the probability can be expressed in terms of $\hbar\omega$ and $\vec{k}$ :
\begin{eqnarray}
& &P_{\vec{k},\hbar\omega}(t_d)=\sum_{\lambda,\lambda '} P^{\lambda,\lambda '}_{\vec{k},\hbar\omega} \nonumber \\
 &= &\sum_{\lambda,\lambda'}2\pi\sigma_{p}^{2}\frac{(d^{cf}\epsilon_1)^2}{\hbar^2}|\alpha_{\lambda'}^{*}(0) ||\alpha_\lambda (0) | \Phi_{\lambda'}(k+K)\Phi_\lambda (k+K)\nonumber\\ & &\times e^{-\frac{\sigma_{p}^{2}}{4\hbar^2}(E_{X_\lambda}-E_{X_{\lambda'}})} e^{-\frac{\sigma_{p}^{2}}{\hbar^2}(\hbar\omega+\frac{\hbar^2(k+K)^2}{2m_h}-\frac{(E_{X_\lambda}+E_{X_{\lambda'}})}{2})^2}\nonumber\\ & &\times e^{\frac{\mathit{i}}{\hbar}( (E_{X_\lambda}-E_{X_{\lambda'}})(t_c-t_p)}
\label{eqmain}
\end{eqnarray}
The probability is expressed as product of two wave functions, oscillating term, Gaussian term that represents the energy conservation rule and Gaussian term with the energy difference. Henceforth it can be represented in a table :\\
\begin{center}
\begin{tabular}{c c c}
$P_{\lambda,\lambda }({\vec{k},\hbar\omega})$&\ldots&$P_{\lambda ,\lambda '}({\vec{k},\hbar\omega})$\\
$\vdots$ & $\ddots$ &$\vdots$\\
$P_{\lambda ',\lambda }({\vec{k},\hbar\omega})$&\ldots & $P_{\lambda ',\lambda '}({\vec{k},\hbar\omega})$
\end{tabular}
\end{center}
The diagonal terms i.e $\lambda=\lambda '$ are the probabilities of exciton in the state $\lambda$. We notice that the oscillating term and the Gaussian term with energy difference are equal to 1. moreover the Gaussian term that reveal the energy conservation rule implies that the spectrum is intense around the $\lambda$ exciton energy. Furthermore the non-diagonal terms constitute the coherent states obtained via the interference of the excitons polarization. We notice that these terms are assified as an oscillation with frequency described by the energy levels difference. Additionnaly the Gaussian term with energy difference presuppose that the signal intensity derived from coherent states is weaker when the two energy levels are far from each other.  And the energy conservation term implies that the signal is crucial around the mean value of the two energies.
\section{Photoemission spectrum of exciton in monolayer $\mathrm{MoS_2}$}
\begin{figure}[H]
\centering
\includegraphics[width=0.45\textwidth , height=0.26\textheight]{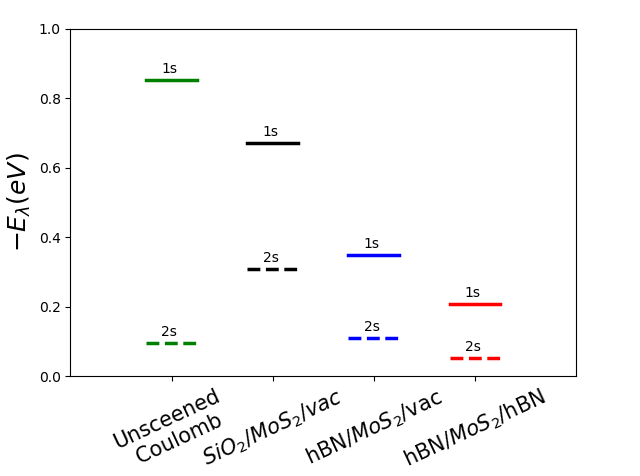}
\caption{Exciton binding energy levels for MoS$_2$ placed on SiO$_2$ ($\kappa = 1.55$), MoS$_2$ placed on hBN ($\kappa =2.75$), hBN encapsulation ($\kappa = 4.5$) and for unscreened Coulomb interaction }
\label{figen}
\end{figure}

Our candidate to study theoretical excitonic signals is the TMD monolayer of MoS$_2$. The electron-hole interaction will be described by Rytova-Keldysh potential for different $\kappa$ and compared to the unscreened Coulomb potential. Recently Chunhao Guo et al. \cite{gap} have shown that the gap energy is modified by the dielectric environment. In this work  we assume that the gap energy of the monolayer will remain unchanged under the effect of the media.
The parameters used to determine exciton eigenvalues and eigenfunctions  as well as the substrate dielectric constants (SiO$_2$, hBN) are listed in the table bellow.
\onecolumngrid\
\begin{center}
\begin{tabular}{c c c c c c c c }
$\mu (m_e)$&$\varepsilon_r$& d(\AA)&W(eV)&a(\AA) &$E_g$(eV)&$\varepsilon_{hBN}$&$\varepsilon_{SiO_2}$ \\
\hline
\hline
0.28 \cite{35} &4.26 \cite{35}&6.15 \cite{1.7}  & 6.12 \cite{37} &3.183 \cite{37} &2.16 \cite{38} &2.1 \cite{ex6} &4.5 \cite{ex6}\\
\hline

\end{tabular}\\
\bigskip
\textbf{Table 1 :} 2H-MoS$_2$ parameters, SiO$_2$ and $hBN$ dielectric constants
\end{center}
\twocolumngrid\
 The convergence of the numerical diagonalisation is obtained for first 10 states but we only  restrict ourselves to 1s and 2s states. For zero magnetic field Molas et al.\cite{mol} have shown that the only the first 5 $\ket{ns}$ states can be observed. Thus we only consider contribution of these states in equation (4) to the eigenstates 1s and 2s. \\
The figure (\ref{figen}) shows the effect of the dielectric environment on energy levels for excitons in a monolayer of $\mathrm{MoS_2}$ :
% and densities of probability in momentum space for excitons in a monolayer of $\mathrm{MoS_2}$ :
%\onecolumngrid\

%\begin{figure}[H]
%\subcaptionbox{\label{figenr1}}{} 
%\subcaptionbox{\label{figphi1}}{}\includegraphics[width=0.3\textwidth , height=0.2\textheight]{../Projet Master/phi1s1.png} 
%\subcaptionbox{\label{figphi2}}{}\includegraphics[width=0.3\textwidth , height=0.2\textheight]{../Projet Master/phi2s1.png} 
%\caption{Exciton binding energy levels (\ref{figenr1}). Density of probability for the 1s state (\ref{figphi1}) and 2s state (\ref{figphi2}) for MoS$_2$ placed on SiO$_2$ ($\kappa = 1.55$), MoS$_2$ placed on hBN ($\kappa =2.75$), hBN encapsulation ($\kappa = 4.5$) and for unscreened Coulomb interaction}
%\label{enden}
%\end{figure}

\twocolumngrid\
\begin{tabular}{c c c}
Substrate & $-E_{1s}(eV)$ & $-E_{2s}(eV)$ \\
\hline
\hline
Unscreened Coulomb & 0.851 & 0.094 \\
SiO$_2$/MoS$_2$/vac & 0.671 & 0.308 \\
hBN/MoS$_2$/vac & 0.348 & 0.1107 \\
hBN/MoS$_2$/hBN & 0.208 & 0.0519\\
\hline

\end{tabular}\\
\bigskip \\
\textbf{Table 2 :} 1s and 2s exciton binding energy for different dielectric environment.
\\

The figure (\ref{figen}) shows that the 1s binding energy for different media decreases from 671 meV for $\kappa =1.55$ to 208 meV for $\kappa =4.5$. The dielectric  environment modify the binding energy of 1s exciton more than 2s exciton. This was explained by Cudazzo et al which implies that when the distance between electron and hole increase i.e larger quantum number, Keldysh potential tend to become unscreened Coulomb potential \cite{cudaz}\cite{cudaz2}. This was also shown by Molas et al. \cite{mol}. \\
 We found the binding energies difference $E_{2s}-E_{1s}$ is smaller for important screening effect $(\kappa =4.5$) than for low screening effect ($\kappa=1.55$). It goes from $\frac{32R_X}{9} (753 meV)$ in Coulomb potential to 363 (156) meV for MoS2 on SiO2 ( hBN encapsulation ) respectively.\\
% Accordingly, figures (\ref{figphi1}) and (\ref{figphi2}) represent the densities of the probability in the relative momentum space. We remark for the same dielectric environment that the density of probability of 1s state is more spread in the momentum space than 2s state because the effective Bohr radius is related to $\kappa$. The larger $\kappa$ gets ( larger radius) the more narrow the density becomes. Meanwhile the magnitude density of probability for 2s state is higher. The results are relevant since the larger the radius gets the broader the wave-function squared becomes in the real space \cite{26}. \\
%To observe the photoemission spectrum resulted from 1s exciton, we set $\hbar\Omega$ to be equal to the 1s exciton energy. the probe energy in vacuum Ultra-violet domain $\hbar\omega_{ph}=12eV$ and the pump/probe temporal width $\sigma_c=\sigma_p=10fs$. Keeping in mind that the smaller the temporal probe width the larger the spectrum along the energy axis. Hence the figure (\ref{figure1s}) illustrates the photoemission spectra derived from equation (\ref{eqmain}) of the 1s exciton for different dielectric environments.
\onecolumngrid\
\begin{figure}[H]
\centering 
\subcaptionbox{\label{fig1s4}}{\includegraphics[scale=0.3]{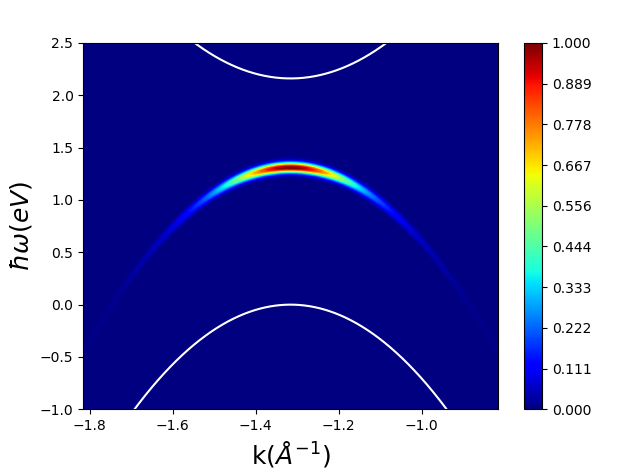}}
\subcaptionbox{\label{fig1s1}}{\includegraphics[scale=0.3]{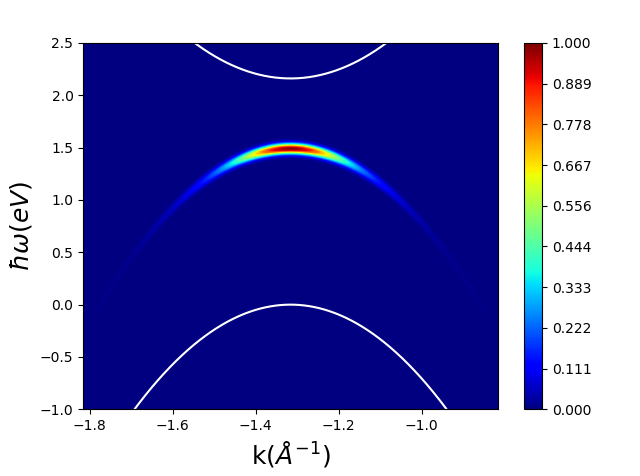}}
\subcaptionbox{\label{fig1s2}}{\includegraphics[scale=0.3]{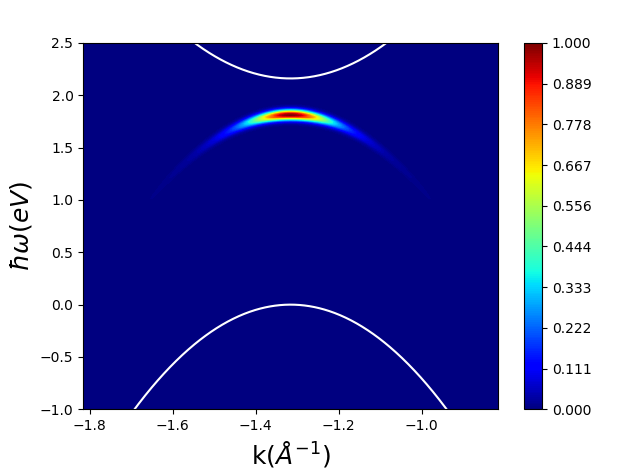}}
\subcaptionbox{\label{fig1s3}}{\includegraphics[scale=0.3]{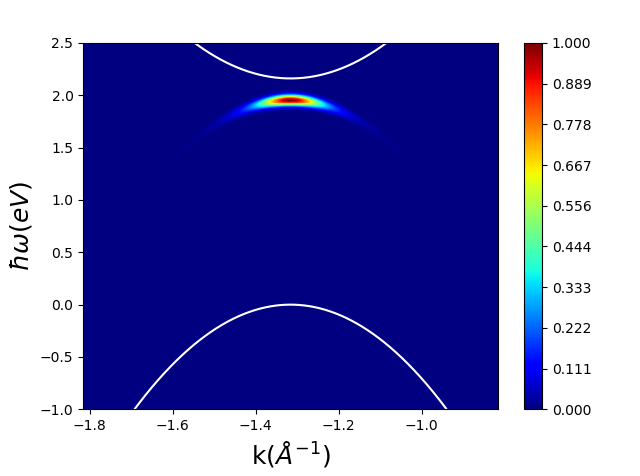}}
\caption{Photoemission spectrum of 1s exciton for unscreened Coulomb potential (\ref{fig1s4}) 1s excitons in MoS$_2$ placed on SiO$_2$ (\ref{fig1s1}) ,hBN (\ref{fig1s2}), between two hBN (\ref{fig1s3})   }
\label{figure1s}
\end{figure}
\twocolumngrid\

\begin{figure}[H]
\centering
\subcaptionbox{\label{figphi1}}{}\includegraphics[width=0.4\textwidth , height=0.2\textheight]{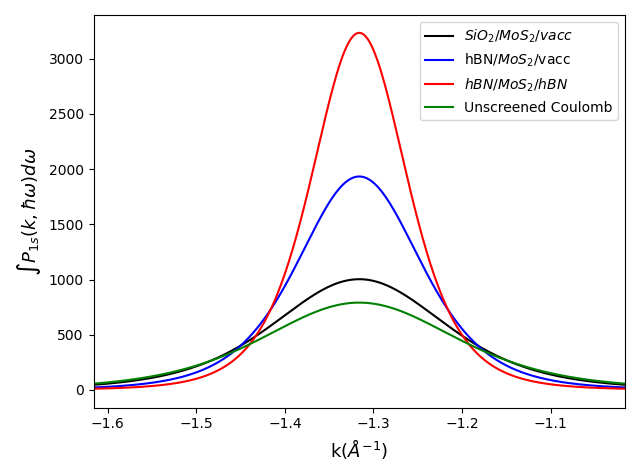}
\subcaptionbox{\label{figphi2}}{}\includegraphics[width=0.4\textwidth , height=0.2\textheight]{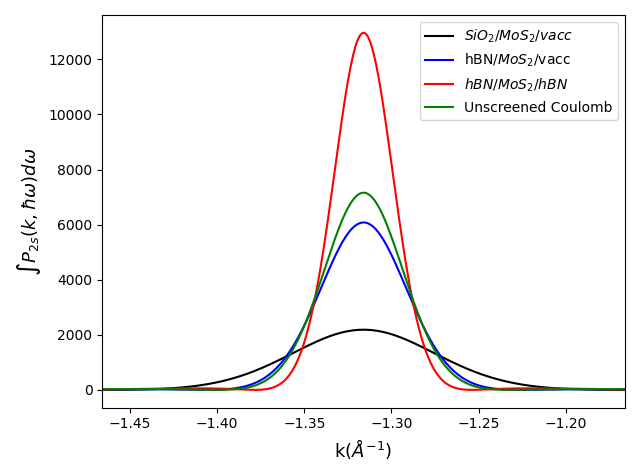} 
\caption{ Density of probability for the 1s state (\ref{figphi1}) and 2s state (\ref{figphi2}) for MoS$_2$ placed on SiO$_2$ ($\kappa = 1.55$), MoS$_2$ placed on hBN ($\kappa =2.75$), hBN encapsulation ($\kappa = 4.5$) and for unscreened Coulomb interaction}
\label{enden}
\end{figure}

The results are similar to experimental work of M.Puppin et al.\cite{40} Mikel K.L et al.\cite{15} and Julien Madéo et al.\cite{12}, which illustrate that the excitons are located between the valence and conduction band within the high symmetry points. The spectrum is intense around the exciton energy and the K valley. However we notice that when $\kappa$ increases, the excitonic signal approaches to the conduction band. \\ For larger $\kappa$ (larger effective Bohr radius) the exciton spectrum becomes narrower along momentum axis which implies that for monolayer embedded between 2 hBN layers only holes around K point can contribute in exciton creation. Unlike when the monolayer placed on SiO$_2$ or hBN where even holes with momentum $\vec{k}_h$ far from K point can contribute to exciton formation which is demonstrated by the valence band-like curve in the figures (\ref{fig1s1})(\ref{fig1s2}). We conclude that for SiO$_2$ and hBN substrates, the Arpes signal reflects clearly the dispersion relation of the valence band.\\
Moreover, the exciton squared wave-function can be deduced via the integration of the probability on the energy axis. In the figure (\ref{enden}) we represent the density of probability of excitons for 1s and 2s states and for different dielectric environments.
\onecolumngrid\
\twocolumngrid\
Accordingly, figures (\ref{figphi1}) and (\ref{figphi2}) represent the densities of the probability in the relative momentum space. We remark for the same dielectric environment that the density of probability of 1s state is more spread in the momentum space than 2s state because the effective Bohr radius is related to $\kappa$. The larger $\kappa$ gets ( larger radius) the more narrow the density becomes. Meanwhile the magnitude density of probability for 2s state is higher. The results are relevant since the larger the radius gets the broader the wave-function squared becomes in the real space \cite{26}. \\
To observe the photoemission spectrum resulted from 1s exciton, we set $\hbar\Omega$ to be equal to the 1s exciton energy. the probe energy in vacuum Ultra-violet domain $\hbar\omega_{ph}=12eV$ and the pump/probe temporal width $\sigma_c=\sigma_p=10fs$. Keeping in mind that the smaller the temporal probe width the larger the spectrum along the energy axis. Hence the figure (\ref{figure1s}) illustrates the photoemission spectra derived from equation (\ref{eqmain}) of the 1s exciton for different dielectric environments.
The momentum integrated $P_{1s}(\vec{k},\hbar\omega)$ is numerically calculated. It provides information about the position and the intensity of the signal on the energy axis.

\begin{figure}[H]
\centering
\includegraphics[scale=0.42]{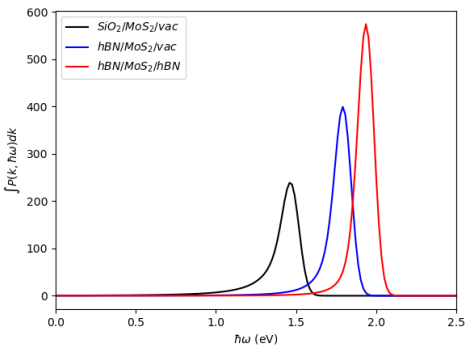} 
\caption{Momentum integrated spectra of excitons in MoS$_2$ for different dielectric environments, the width of the curve is controlled by the temporal probe width $\sigma_p$}
\end{figure}

the peaks are located respectively in 1.489eV for SiO$_2$/MoS$_2$/vac, 1.852eV for hBN/MoS$_2$/vac and 1.952eV for hBN/MoS$_2$/hBN. Furthermore the intensity is more important when the monolayer is encapsulated between two hBN layers because of its high transparency that preserves excitonic signals, also high dielectric constant and large band gap which prevent exciton quenching \cite{15}. \\
A resonant excitation is convenient to observe 1s excitonic signals. However since the density of probability of 2s state is larger than 1s state as discussed in section 2, it is trivial that the signal of 2s exciton will be more intense. Thus to observe the coherent terms, it is important to choose the pump energy $\hbar\Omega$ in way that we obtain equal contribution from both exitons. Therefore we numerically calculate the contrast between the two excitons signals :
\begin{equation}
C=\frac{|P_{1s}^{max}-P_{2s}^{max}|}{P_{1s}^{max}+P_{2s}^{max}}
\end{equation}
Where $P_{1s}^{max}(P_{2s}^{max})$ correspond to the maximum value of the excitonic 1s (2s) signals. $\hbar\Omega$ is set in a way that we obtain a minimum contrast $P_{1s}^{max}\simeq P_{2s}^{max}$. Thus we set $\hbar\Omega=1.66eV$ for MoS$_2$ placed on SiO$_2$, $\hbar\Omega=1.91eV$ for MoS$_2$ placed on hBN and  $\hbar\Omega=2.04eV$ for MoS$_2$ encapsulated between two hBN layers.\\ 
The dependency on the time delay in non-diagonal terms  will lead to a quantum beats. We first investigate excitonic signals for MoS$_2$ placed on hBN. 
\onecolumngrid\
\begin{figure}[H]
\centering
\subcaptionbox{\label{cohfighbn1}}{\includegraphics[scale=0.29]{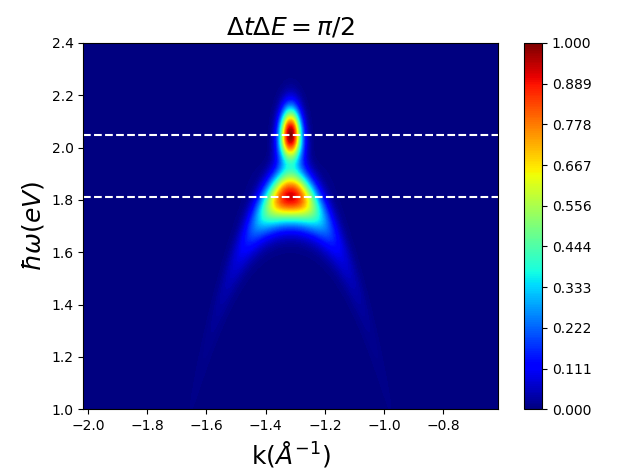} }
\subcaptionbox{\label{cohfighbn2}}{\includegraphics[scale=0.29]{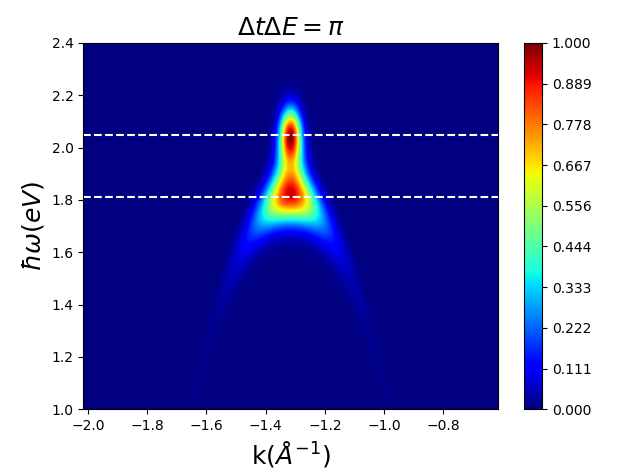} }
\subcaptionbox{\label{cohfighbn3}}{\includegraphics[scale=0.29]{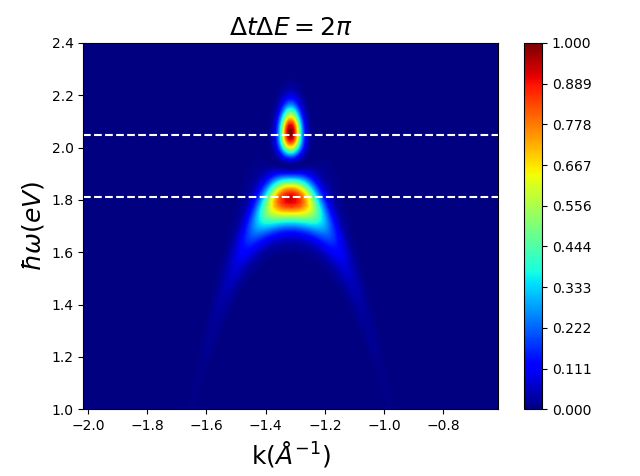}}
\caption{Photoemission spectrum of coherent excitons  in MoS$_2$ placed on hBN. Noting that the 1s and 2s excitonic spectra are time invariant. However the interference spectrum is oscillating in time}
\label{fihcohhbn}
\end{figure}
\twocolumngrid\
The figure (\ref{fihcohhbn}) the MoS$_2$ is placed on hBN. We capture the photoemission spectra for different time delays $t_b=t_c-t_p$. We have $t_b=\frac{\pi\hbar}{2|E_{X_{1s}}-E_{X_{2s}}|}$ (\ref{cohfighbn1}), $t_b=\frac{\pi\hbar}{|E_{X_{1s}}-E_{X_{2s}}|}$(\ref{cohfighbn2}) and $t_b=\frac{2\pi\hbar}{|E_{X_{1s}}-E_{X_{2s}}|}$ (\ref{cohfighbn1}). The 2s spectra is narrower than 1s signal along the momentum axis and close to the conduction band along the energy axis. Since $P_{\lambda}(\vec{k},\hbar\omega)\propto |\Phi_{\lambda}(\vec{k})|^2$, the figure (\ref{enden}) showed that 2s exciton denisty is less spread in k space than 1s exciton density. This suggests also that only holes with momentum near the valley K can contribute to 2s exciton formation.\\

The beats signals are formed in the region between the two excitons spectrum. As time progresses, the polarization interference displays periodic fluctuations. We notice that the interference represents a maxima for  $\Delta E \Delta t=\pi\hbar$ ($t_c-t_p=8.8fs$), minima for $\Delta E \Delta t=2\pi\hbar$ ($t_c-t_p=17.6fs$) and for $\Delta E \Delta t=\frac{\pi}{2}\hbar$ ($t_c-t_p=4.4fs$), $P_{1s,2s}$ and $P_{2s,1s}$ cancel each other. \\
In order to observe the beat evolution through time, we numerically calculate the momentum integrated probability for excitons in MoS$_2$ placed on hBN for different value of probe temporal width $\sigma_p$ and we set $\sigma_c=10fs$.
\onecolumngrid\
\begin{figure}[H]
\centering
\subcaptionbox{\label{cohfighbn4}}{\includegraphics[scale=0.29]{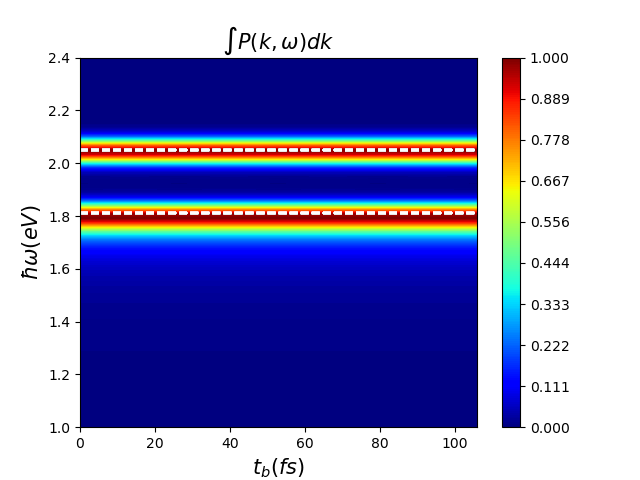} }
\subcaptionbox{\label{cohfighbn5}}{\includegraphics[scale=0.29]{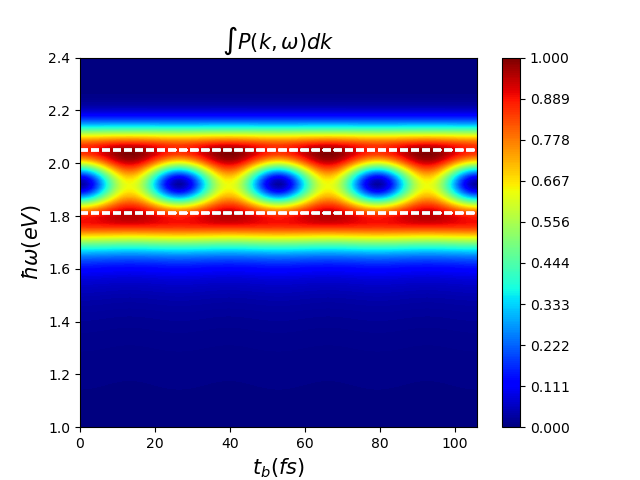}  }
\subcaptionbox{\label{cohfighbn6}}{\includegraphics[scale=0.29]{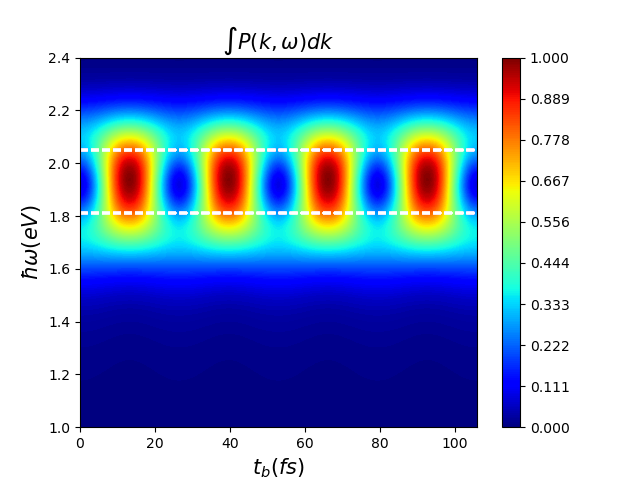} }
\caption{momentum integrated probability of coherent excitons  in MoS$_2$ placed on hBN for $\sigma_p=15fs$ (\ref{cohfighbn4}), $\sigma_p=7fs$ (\ref{cohfighbn5}) and $\sigma_p=4fs$ (\ref{cohfig6})} 
\label{fihcohhbn1}
\end{figure}
\twocolumngrid\
 The probe temporal width $\sigma_p$ is an important parameter to observe the beats. Therefore for MoS$_2$ on hBN substrate we notice when $\sigma_p$ is large, the beat signals are suppressed. On the other side when the pulse temporal width is low we obtain an overlap between the two signals arising from 1s and 2s excitons. Therefore we set the temporal width $\sigma_p=7fs$ for MoS$_2$ placed on SiO$_2$ and hBN, $\sigma_p=10fs$ for monolayer encapsulated between two hBN layers and $\sigma_p=4fs$ for unscreened Coulomb potential :
 \onecolumngrid\
\begin{figure}[H]
\centering
\subcaptionbox{\label{cohfig4}}{\includegraphics[scale=0.32]{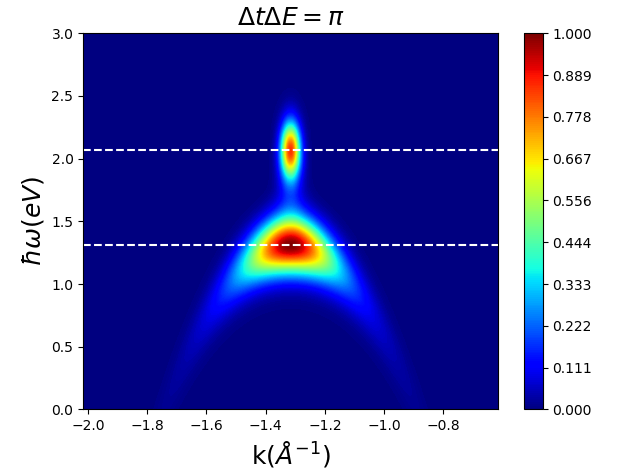}}
\subcaptionbox{\label{cohfig1}}{\includegraphics[scale=0.32]{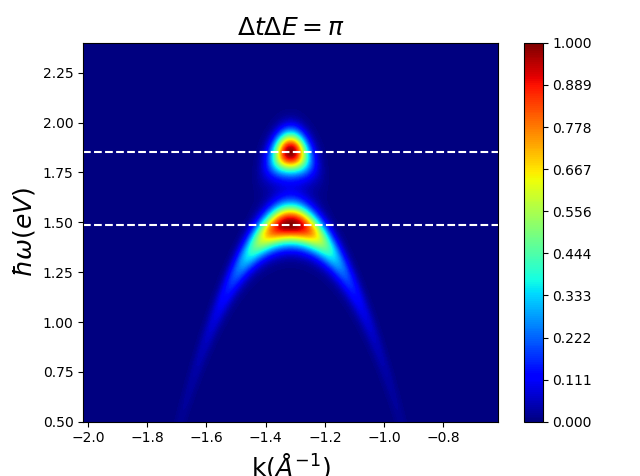} }
\subcaptionbox{\label{cohfig2}}{\includegraphics[scale=0.32]{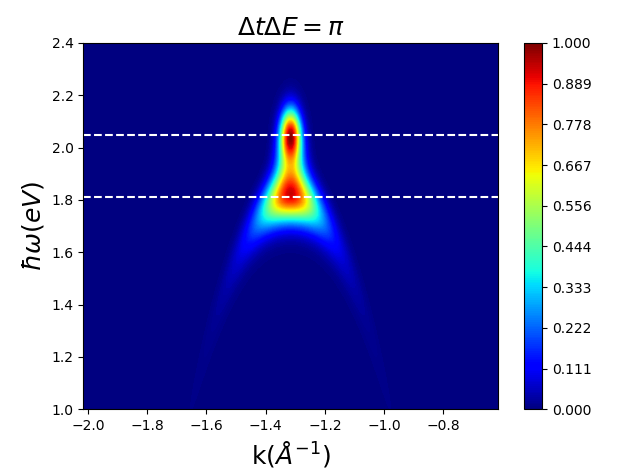} }
\subcaptionbox{\label{cohfig3}}{\includegraphics[scale=0.32]{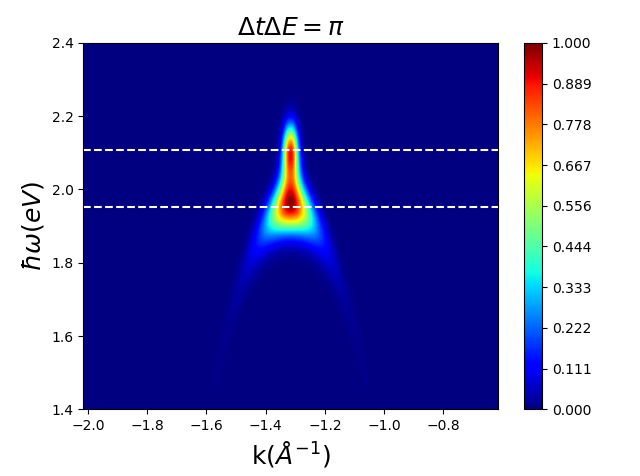}}
\caption{Photoemission spectrum of coherent excitons for unscreened Coulomb potential
 (\ref{cohfig4}), for MoS$_2$ placed on SiO$_2$ (\ref{cohfig1}), placed on hBN (\ref{cohfig2}), MoS$_2$ encapsulated between two hBN layers (\ref{cohfig3})}
\label{fihcoh}
\end{figure}
\twocolumngrid\
The figure (\ref{fihcoh}) shows that the two excitons are formed simultaneously. The coherent exciton signals appear in the region between two energy levels. Therefore the two excitons are behaving like a single quasi particle. Moreover for hBN/MoS$_2$/hBN (Figure (\ref{cohfig3})) the two excitons signals are closer than for SiO$_2$/MoS$_2$/vac (Figure (\ref{cohfig1})) and hBN/MoS$_2$/vac (Figure (\ref{cohfig2})) which enhance the quantum interference hBN encapsulation. As result, it yields to obtain different oscillation frequency for every substrate. To gain a deeper understanding about these beats and the periodic oscillation we plot the momentum integrated probability.
\onecolumngrid\
\begin{figure}[H]
\centering
\subcaptionbox{\label{cohfig6}}{\includegraphics[scale=0.35]{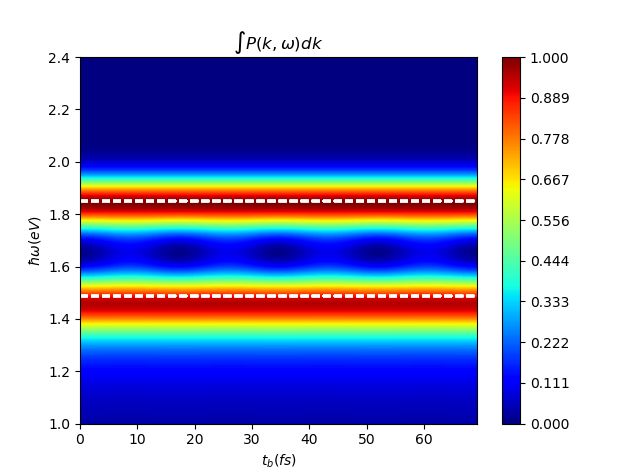} }
\subcaptionbox{\label{cohfig7}}{\includegraphics[scale=0.35]{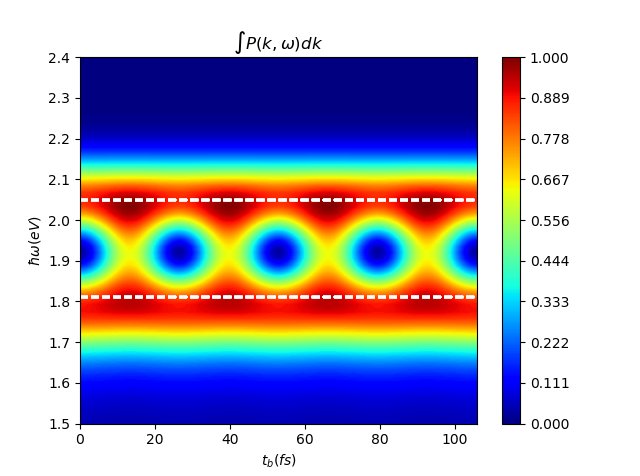} }
\subcaptionbox{\label{cohfig8}}{\includegraphics[scale=0.35]{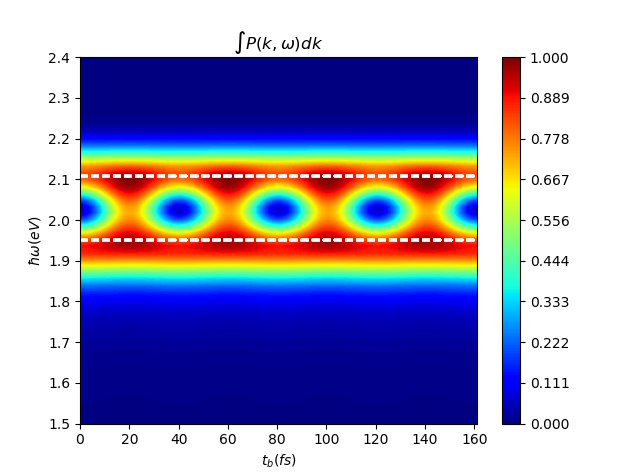}}
\subcaptionbox{\label{cohfig9}}{\includegraphics[scale=0.35]{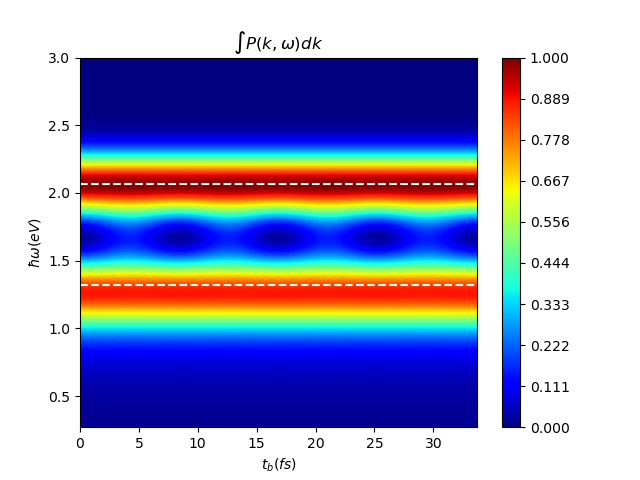}  }
\caption{Momentum integrate of Photoemission spectrum of coherent excitons  in MoS$_2$ placed of SiO$_2$ (\ref{cohfig6}), placed on hBN (\ref{cohfig7}), MoS$_2$ encapsulated between two hBN layers (\ref{cohfig8}) and unscreened Coulomb potential (\ref{cohfig9})}
\label{fihcoh1}
\end{figure}
\twocolumngrid\
We notice in the figure (\ref{fihcoh1}) that the beat effect is important for hBN substrate and hBN encapsulation. However we notice fo SiO$_2$ substrate the beat is slightly observed because of the important energy difference.
The oscillation period $T=\frac{2\pi\hbar}{|E_{X_{1s}}-E_{X_{2s}}|}$ is influenced by the environment. We notice that when $\kappa$ increases, the period of oscillation becomes larger which make encapsulating MoS$_2$ between two hBN layer appropriate to observe this phenomena since $T=26.6fs$. However for for MoS$_2$ placed on hBN we have $T=17.6fs$ which yields to multiple oscillation before electron recombination. 
Moreover the oscillation period of 2s-3s excitons is larger since the two levels are closer. For MoS$_2$ placed on hBN we illustrate the momentum integrated Photoemission spectra in the figure bellow :
\begin{figure}[H]
\centering
\includegraphics[width=0.4\textwidth , height=0.25\textheight]{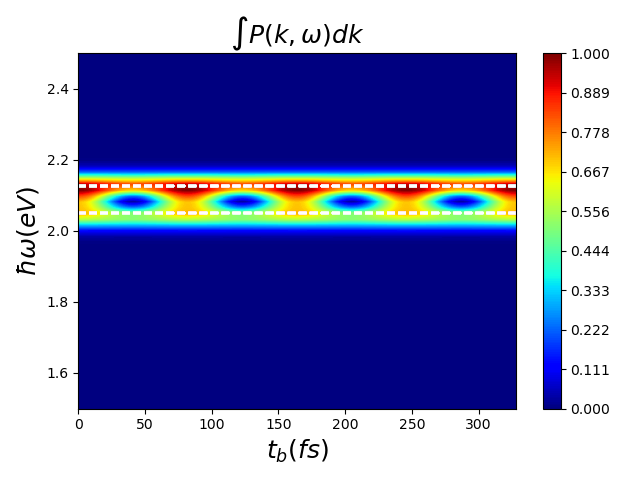} 
\caption{The momentum integrated spectra of coherent 2s-3s exciton in MoS$_2$ placed on hBN. we assume that the 3s state is not affected by the dielectric environment and the interaction between electron and hole is unscreened Coulombian. to obtain this coherent state we set $\sigma_c=\sigma_p=15fs$ and $\hbar\Omega=2.08eV$}
\label{3s2s}
\end{figure} 
In the Figure (\ref{3s2s}) we notice that the period of oscillation is around $T=51fs$. Thus we conclude that it is possible to control this oscillation by choosing the two energy levels.
\section{Conclusion}
Ultimately, Unlike the bulk semi-conductors, the dielectric environment have an impact on the binding energy and the probability of presence of 2D excitons. As a result it yields to changing the photoemission signals of excitons for different substrates. Momentum integrated spectra have demonstrated that the intensity is more important for 2D TMD materials placed on substrate with large dielectric constant. Furthermore, using the adequate parameters such as the pump energy will  lead to the observation of coherent excitons. The interference of the excitons polarization display oscillations where the frequency is determined by the energy difference. This interference created quantum beats between 1s and 2s excitons. The dielectric environment plays a crucial role in the beats period. for high dielectric constants ($\kappa =4.5)$ we observe a decrease of the oscillation frequency.
% These quantum beats can't oscillate forever since the excitons in ML TMDs have a lifetime around 0.1ps$\sim$1ps. However in TMDs heterostrcuture, a significant increase in excitons lifetime reaching 30ns have been demonstrated \cite{life} which make these beats last longer.
%\section*{Acknowledgement}
%We would like to record our appreciation to Ermin Malik, Alexander F.Kemper and Maciej Molas for their  constructive criticisms and valuable
%comments.
\bibliographystyle{unsrt}
\bibliography{biblioart}
\appendix
\section{ the time dependent wave function after the pump}
In the moment $t=t_0$ the system is in the ground state $\ket{0}$. Afterwards the pump acts at the moment $t_c-\sigma_c$ and ends at $t_c+\sigma_c$. Hence the dynamic of the state is described using Heisenberg description, the state $\ket{\Psi(t)}$ :
\begin{eqnarray}
& &\ket{\Psi(t)}=U_X(t,t_0)\ket{0} \nonumber \\
& & = e^{-\frac{\mathit{i}}{\hbar}\int_{t_0}^{t}(H_X+E_0+H_{vc}(\tau))d\tau}\ket{0} \nonumber \\
&& = \prod_\lambda e^{-\frac{\mathit{i}}{\hbar}\int_{t_0}^{t}(H_{X_\lambda}+E_0)d\tau} D_\lambda (\alpha_\lambda)\ket{0}
\end{eqnarray} 
 $D_\lambda (\alpha_\lambda)=e^{\alpha_\lambda X^{\dag}_{\lambda}-\alpha^{*}_{\lambda}X_{\lambda}}$ is analogue to the the displacement operator established by R.J Glaubert to describe coherent systems \cite{30}\cite{31}. Furthermore $\alpha_\lambda =\frac{\mathit{i}}{\hbar}\int_{t_c-\sigma_c}^{t_c+\sigma_c}\epsilon_{vc}(\tau)d_\lambda e^{\mathit{i}\frac{E_{X_\lambda}}{\hbar}\tau}d\tau$ is the complex number  but since the pump pulse is zero far from [$t_c-\sigma_c,t_c+\sigma_c$] we can extend the integral to ]$-\infty,+\infty$[. We pose $\tau '=\tau-t_c$ in $\vec{\epsilon}_{vc}(\tau)$ expression and we neglect therm with $e^{-\frac{\sigma_{c}^{2}}{2\hbar^2}(E_{X_\lambda}+\hbar\Omega)^2}$ because it.s very small compared to the term with $e^{\frac{-\sigma_{c}^{2}}{2\hbar^2}(E_{X_\lambda}-\hbar\Omega)^2}$  :
\begin{equation}
\alpha_\lambda (0) \simeq\mathit{i}\sqrt{\pi}\frac{\sigma_c}{\hbar}\epsilon_{0}d_\lambda e^{-\frac{\sigma_{c}^{2}}{2\hbar^2}(E_{X_\lambda}-\hbar\Omega)^2}e^{\frac{\mathit{i}}{\hbar}E_{X_\lambda}t_c}
\end{equation}
 The complex number $\alpha_\lambda (0) $ depends on the transition matrix element, the electromagnetic field, a phase factor $e^{\frac{\mathit{i}}{\hbar}E_{X_\lambda}t_c}$ as well as the probe energy and the $\lambda$ exciton energy difference. Thus $|\alpha_\lambda (0)|$ tend to 0 if we choose $\hbar\Omega$ far from the exciton energy $E_{X_\lambda}$.
The action of the displacement operator on the ground state $\ket{0}$ is creating a coherent state $\ket{\alpha_\lambda}$ with eigenvalues $\alpha_\lambda (t)= \alpha_\lambda (0)e^{-\mathit{i}\frac{E_{X_\lambda}}{\hbar}t}$. This allows us to write the time dependent state of the system as \cite{32} :
\begin{equation}
\ket{\Psi(t)}= \prod_\lambda e^{-\frac{\mathit{i}}{\hbar}(E_0+H_{X_\lambda})(t-t_0)} e^{-\frac{|\alpha_\lambda (0)|^2}{2}}e^{\alpha_\lambda (t) X^{\dag}_{\lambda}}\ket{0}
\end{equation}
\section{The time dependent wave-function after the probe}
the probe pulse will take a place at $t_p-\sigma_p$ and ends at $t_p+\sigma_p$. Thus the evolution of the system after the probe is given by :
\begin{eqnarray}
& & \ket{\Psi^F(t)}=U_p(t,t_0)\ket{\Psi(t)}\nonumber\\
& & =e^{-\frac{\mathit{i}}{\hbar}\int_{t_0}^{t}H_fd\tau}e^{-\frac{\mathit{i}}{\hbar}\int_{t_0}^{t}H_{cf}(\tau)d\tau}\ket{\Psi(t)}\nonumber\\
\end{eqnarray}
Supposing that the probe pulse magnitude is low. Hence we treat $H_{cf}(\tau)$ as perturbation which will allow us to linearise $U_p(t,t_0)$ to obtain \cite{34} :
\begin{eqnarray}
&& \ket{\Psi^F(t)}\simeq   e^{-\frac{\mathit{i}}{\hbar}H^f(t-t_0)}\nonumber\\
& & \times(1-\frac{\mathit{i}}{\hbar}\int_{t_0}^{t}U_X(t,\tau)H_{cf}(\tau)U_X(\tau ,t_0)d\tau)\ket{0}\nonumber\\
\end{eqnarray}
We note that $a_f(\vec{k}_e)\ket{\Psi(t)}=0$ and the momentum conservation rule implies that $\vec{k}_{e}-\vec{k'}-\vec{K}=\vec{0}$ i.e the electron undergoes direct transition from conduction band to vacuum. Thus the first order term have the following expression.   :
\begin{eqnarray}
& & U_{p}^{(1)}(t,t_0)=-\frac{\mathit{i}}{\hbar}\int_{t_0}^{t}\tilde{U}(t,\tau_1)\sum_{\vec{k'}}\hbar\Omega_{R}^{cf}(\tau_1)\nonumber\\ & &\times a_{f}^{\dag}(\vec{k'}+\vec{K})a_c(\vec{k'}+\vec{K})\tilde{U}(\tau_1,t_0)d\tau_1
\end{eqnarray} 
With $\tilde{U}(t,\tau_1)=e^{-\frac{\mathit{i}}{\hbar}H^f(t-\tau_1})U_X(t,\tau_1)$. First step is to evaluate $a_c(\vec{k'})\tilde{U}(\tau_1,t_0)$ using the following commutation relation [A,f(B)]=[A,B]$\frac{d f(B)}{d B}$. the momentum conservation rule presuppose that $\vec{k'}+\vec{K}=\vec{K}+\vec{k}$. As consequence we obtain :
\begin{eqnarray}
 a_c(\vec{k'})\tilde{U}(\tau_1,t_0)&=&(\sum_\lambda \alpha_\lambda (\tau_1) \Phi_\lambda(\vec{K}+\vec{k}))a_v(\vec{K}+\vec{k})\nonumber\\
& &\times e^{-\frac{\mathit{i}}{\hbar}(\frac{\hbar^2(\vec{K}+\vec{k})^2}{2m_e}+W+H_{X}+E_0)(\tau_1-t_0)}\nonumber\\ & &\times \prod_\lambda e^{-\frac{|\alpha_\lambda|^2}{2}}e^{\alpha_\lambda (t) X^{\dag}_{\lambda}}
\end{eqnarray}
The probability to detect a photoemitted electron in the state $\ket{f}$ resulted from excitonic states is :
\begin{eqnarray}
& & P(t_d)=\left|\left|\bra{f}\ket{\Psi^F(t)}\right|\right|^2 \nonumber\\
& &=\frac{(d^{cf}\epsilon_1)^2}{\hbar^2}\int_{t_0}^{t}d\tau_2\int_{t_0}^{t}d\tau_1 e^{-\frac{(\tau_2-t_p)^2}{2\sigma_{p}^{2}}}e^{-\frac{(\tau_1-t_p)^2}{2\sigma_{p}^{2}}}\nonumber\\
& &\times e^{\frac{\mathit{i}}{\hbar}(\hbar\omega-E_v(\vec{K}+\vec{k})-\sum_\lambda E_{X_\lambda} )(\tau_2-\tau_1)}\nonumber\\
& &\times(\sum_{\lambda'}\alpha^{*}_{\lambda'}(\tau_2)\Phi^{*}_{\lambda'}(\vec{K}+\vec{k})) (\sum_{\lambda}\alpha_{\lambda}(\tau_1)\Phi_{\lambda}(\vec{K}+\vec{k}))\nonumber\\
\end{eqnarray}
with $\hbar\omega=\frac{\hbar^2k^2}{2m_e}+W-\hbar\omega_{ph}$ is the energy difference of the ejected electron. Bearing in mind that when the pump take place and the electron is photoemitted leaving a whole in the solid band, the energy of the system becomes $E_0-E_v(\vec{K}+\vec{k})$ where $E_v(\vec{K}+\vec{k})=-\frac{\hbar^2k^2}{2m_h}$ is the energy of the electron in the valence band. We expand the integrals of $\tau_1$ and $\tau_2$ from [$t_0,t]$ to ]$-\infty,+\infty$[.
\end{document}